\documentclass[11pt]{article}
\usepackage{graphics,cite,amssymb,epsfig,float}
\usepackage[usenames,dvips]{color}
\usepackage{amsmath, mathbbol}

\usepackage{multirow}
\begin{document}

\setcounter{footnote}{0}
\vspace*{-2.cm}
\begin{flushright}
PCCF RI 14-02\\
LPT Orsay 14-08 \\
CFTP/14-001
\vspace*{2mm}
%\today								
\end{flushright}
\begin{center}
\vspace*{5mm}

\vspace{1cm}
{\Large\bf 
Lepton flavour violation at high energies: the LHC and \\\vspace{.2cm}
a Linear Collider}\\
\vspace{.7cm}

{\bf  A.M. Teixeira$^{a}$\footnote{Proceedings of 
the ``Linear Collider Workshop - LC13'', ECT$^*$ Trento, 16 - 
20 September 2013.}, 
A. Abada$^{b}$, A. J. R. Figueiredo$^{a,c}$ and
  J. C. Rom\~ao$^{c}$ }

\vspace*{.5cm} 

$^{a}$ Laboratoire de Physique Corpusculaire, CNRS/IN2P3 -- UMR 6533,\\ 
Campus des C\'ezeaux, 24 Av. des Landais, F-63171 Aubi\`ere Cedex, France
\vspace*{.2cm}

$^{b}$ Laboratoire de Physique Th\'eorique, CNRS -- UMR 8627, \\
Universit\'e de Paris-Sud 11, F-91405 Orsay Cedex, France
\vspace*{.2cm} 

$^{c}$ Centro de F\'{\i}sica Te\'orica de Part\'{\i}culas, CFTP, 
Instituto Superior T\'ecnico, \\
Universidade de Lisboa, Av. Rovisco Pais 1, 
1049-001 Lisboa, Portugal

\end{center}

\vspace*{5mm} 

\begin{abstract}
We discuss several manifestations 
of charged lepton flavour violation at high energies.
Focusing on a supersymmetric type I seesaw, considering constrained
and semi-constrained supersymmetry breaking scenarios, we analyse
different observables, both at the LHC and at a future Linear
Collider. 
We further discuss how the synergy between low- and high-energy 
observables can shed some light on
the underlying mechanism of lepton flavour violation.

\vspace*{5mm}

\noindent
{\small {\sc PACS} \texttt{14.60.St}{ Non-standard-model neutrinos}}\\
{\small {\sc PACS} \texttt{12.60.Jv}{ Supersymmetric models}}\\
{\small {\sc PACS} \texttt{13.66.Hk}{ 
Production of non-standard model particles 
in $e^+e^-$ interactions}}\\
{\small {\sc PACS} \texttt{13.35.Bv}{ Decays of muons}}

\end{abstract}

%% \PACSes{\PACSit{14.60.St}{Non-standard-model neutrinos}
%% \PACSit{12.60.Jv}{Supersymmetric models}
%% \PACSit{13.66.Hk}{Production of non-standard model particles 
%% in $e^+e^-$ interactions}
%% \PACSit{13.35.Bv}{Decays of muons}}

\vspace*{5mm}

\section{Motivation}
The observation of neutrino oscillations has provided the first clear
evidence of new physics beyond the Standard Model (SM). In the
SM lepton flavour is strictly conserved; by themselves, neutrino
oscillations signal the violation of neutral lepton flavour, and open
the door to general scenarios where, in analogy to the quark
sector, both neutral and charged lepton flavours are no longer
conserved. 
Neutrino phenomena can be trivially accommodated 
by minimal extensions of the SM,
where (Dirac) neutrino mass terms are put by hand, and flavour mixing
in the leptonic sector is parametrised by 
the $U_\mathrm{PMNS}$ (the SM$_{m_\nu}$). 
However, in the context of such a
simple framework, the contributions to charged lepton flavour
violation (cLFV) observables, such as $\mu \to e \gamma$, are expected
to be very small, beyond any experimental reach. This implies that the
observation of any cLFV manifestation would indisputably reveal the
presence of New Physics (NP) other than the SM$_{m_\nu}$.  

Among many appealing NP models accommodating neutrino data (for a
brief overview see~\cite{Abada:2011rg}), here we consider the 
hypothesis of a supersymmetric (SUSY) type I seesaw, which can open
the door to a large number of
cLFV observables at/below the TeV scale. 
These observables can be searched for 
in low-energy, high intensity facilities or in high-energy colliders
as the LHC or a future Linear Collider (LC).  
We focus on the prospects of a SUSY seesaw concerning 
cLFV signals at high-energies:  
we first re-evaluate the
prospects for observing cLFV observables at the LHC (such as new edges
in dilepton mass distributions and flavoured
slepton mass differences), and whether the type-I SUSY seesaw can
still be probed via the synergy between high- and low-energy cLFV
observables\cite{Abada:2010kj,Figueiredo:2013tea}.
We then address the potential of a Linear Collider regarding lepton
flavour violation, in particular 
$e\,\mu +$ missing energy final states, 
arising from $e^+ e^-$ and $e^- e^-$ 
collisions~\cite{Abada:2012re}. We address the potential 
background from SM and SUSY charged
currents, also exploring the possibility of electron and positron beam
polarisation. We finally discuss a potential ``golden channel''for
cLFV at a LC: $e^- e^- \to \mu^-\mu^- + E_\text{miss}$.

\section{Theoretical framework: the SUSY seesaw}\label{susyseesaw}
We consider a supersymmetric version of a type I seesaw, where 3 
right-handed (RH)
neutrino superfields are added to the Minimial Supersymmetric SM 
(MSSM) particle content. 
In our analysis we assume a flavour-blind SUSY breaking
mechanism where the soft breaking parameters satisfy universality
conditions at some high-energy scale ($M_\mathrm{GUT} \sim 10^{16}$
GeV). We consider cases of strict universality, i.e. an embedding
onto the constrained MSSM (cMSSM), as well as 
semi-constrained frameworks, where one breaks strict universality for
squark, slepton and Higgs soft-breaking terms at $M_\mathrm{GUT}$ (but
still preserving flavour universality), also assuming that gluino and
electroweak (EW)
gaugino masses are independent (for a detailed discussion, 
see~\cite{Figueiredo:2013tea}). 

After EW symmetry breaking, the light neutrino masses are
given by the seesaw relation, $m_\nu \simeq -v^2_2 {Y^\nu}^T M^{-1}_R Y^\nu$
which in turn suggests a convenient parametrisation for the neutrino
Yukawa couplings~\cite{Casas:2001sr},
$Y^\nu\, {v_2}= i\, \sqrt{M^\text{diag}_R} \, R \, 
\sqrt{m^\text{diag}_\nu} \, {U_\text{PMNS}}^\dagger $,
where $v_2$ is one of 
the vacuum expectation values of the neutral Higgs, 
$U_\text{PMNS}$ is the leptonic mixing matrix and 
$R$ is a complex orthogonal matrix, parameterised in terms of three
complex angles, 
that encodes additional mixings involving the RH neutrinos; 
$m^\text{diag}_\nu$ and $M^\text{diag}_R$ 
denote the (diagonal) light and heavy neutrino mass
matrices.

In the SUSY seesaw, flavour mixing in the slepton sector is
radiatively induced due to the non-trivial structure of the neutrino
Yukawa couplings~\cite{Borzumati:1986qx}. 
Even in the case of universal SUSY breaking
terms, renormalisation-group 
running will give rise to flavour-violating entries in the
slepton masses. The misalignment of  
flavour and physical slepton eigenstates leads to LFV neutral and
charged lepton-slepton interactions, and will be manifest in
contributions to numerous cLFV observables. 
Having posited a unique source of LFV ($Y^\nu$) implies that all
observables will exhibit a certain degree of correlation, 
which in turn might
allow to indirectly probe the high-scale seesaw hypothesis.

At low-energies, virtual SUSY particles mediate rare transitions and
decays. In our work, we mostly focus on
raditive muon decays, whose recent bound from MEG is 
BR$(\mu \to e \gamma) \lesssim 5.7 \times 10^{-13}$~\cite{Adam:2013mnn} .

High-energy colliders (such as the LHC or a future LC) provide direct
access to the slepton sector; provided the centre of mass energy is
sufficiently large, one can study on-shell sleptons, either directly
produced (as in a LC), or then arising from decay chains of directly
produced coloured states or EW-inos (as at the LHC). It is important
to notice that LFV in charged current interactions (e.g. 
$\chi^\pm - \tilde \ell_i - \nu_j$ interactions) 
can arise as a simple consequence of
an ad-hoc implementation of leptonic mixing, and does not signal 
any new physics beyond the SM$_{m_\nu}$ (in this case the MSSM$_{m_\nu}$). On
the other hand, LFV in neutral currents indisputably signals such new
physics (e..g., the type I SUSY seesaw). 

In the following sections we illustrate some cases of cLFV at
high-energies,  mostly focusing on flavour violation in the
(s)electron-(s)muon sector\footnote{For a discussion of 
flavour violating final states with a $\tau$ lepton, 
in particular $\mu \tau + E^T_\mathrm{miss}$,
  see M. E. G\'omez, these proceedings.}.

\section{cLFV at the LHC}
To illustrate cLFV observables at the LHC, we consider slepton
production from wino-like neutralino decays. 
The shape of the dilepton mass distribution from
$\chi_2^0 \to \tilde \ell \ell \to \ell \ell \chi_1^0$ decays allows
to extract information on the slepton spectrum; in particular, the
position of the edges allows to determine the slepton mass, while the
number of edges translates the different number of 
sleptons participating in the decay. 

\begin{figure}
\begin{tabular}{ccc}
\raisebox{3mm}{\epsfig{file=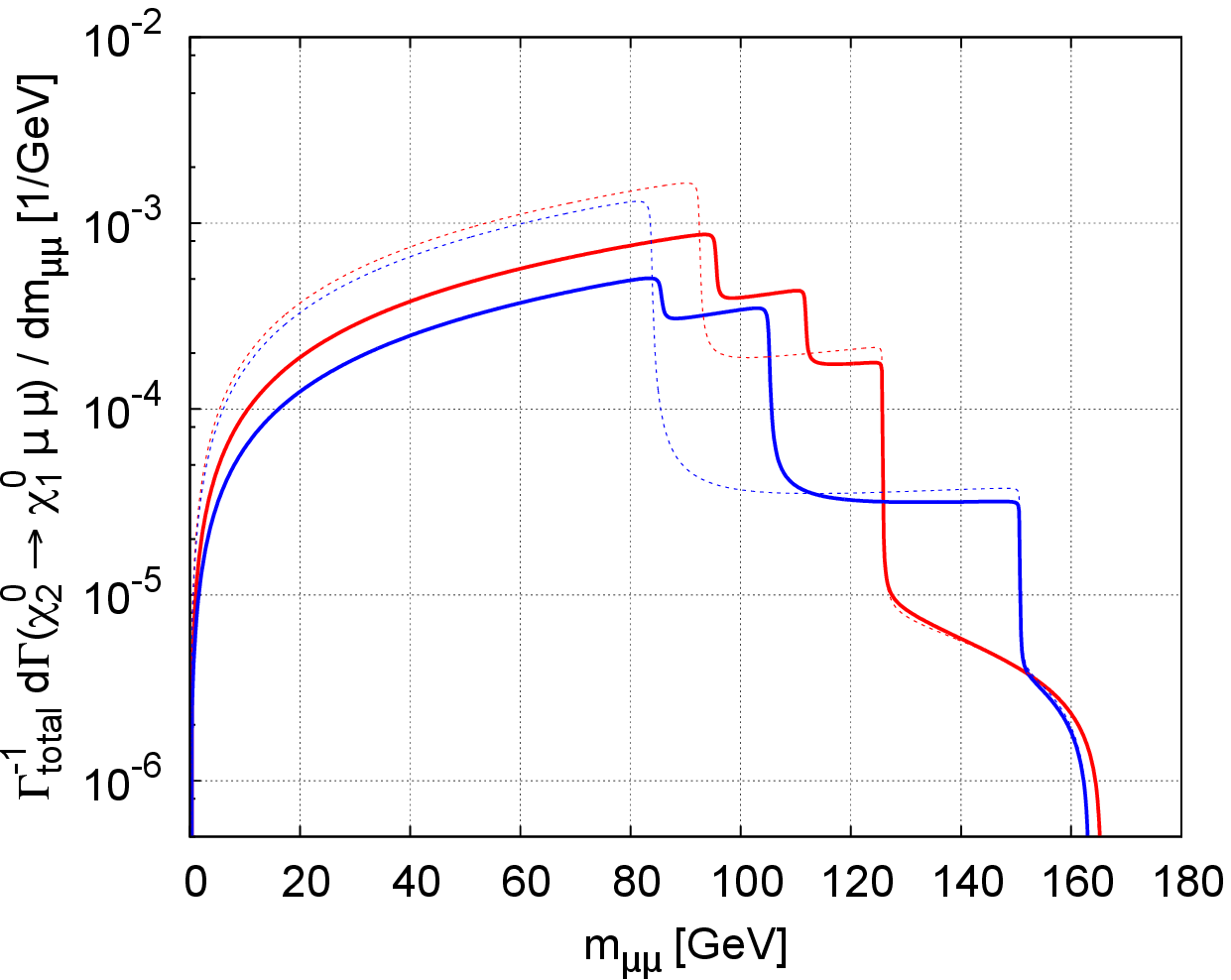, width=45mm}}
\hspace*{-1mm}&\hspace*{-1mm}
\epsfig{file=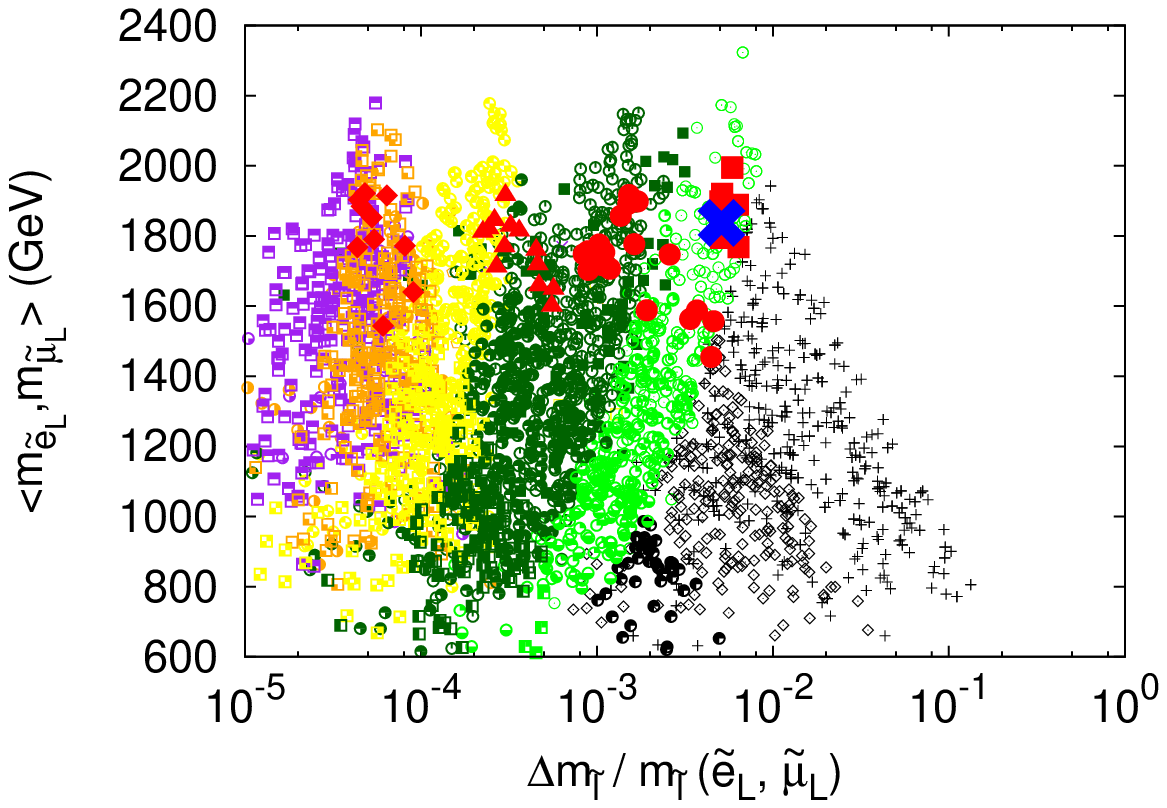, width=55mm}
\hspace*{-1mm}&\hspace*{-3mm}
\epsfig{file=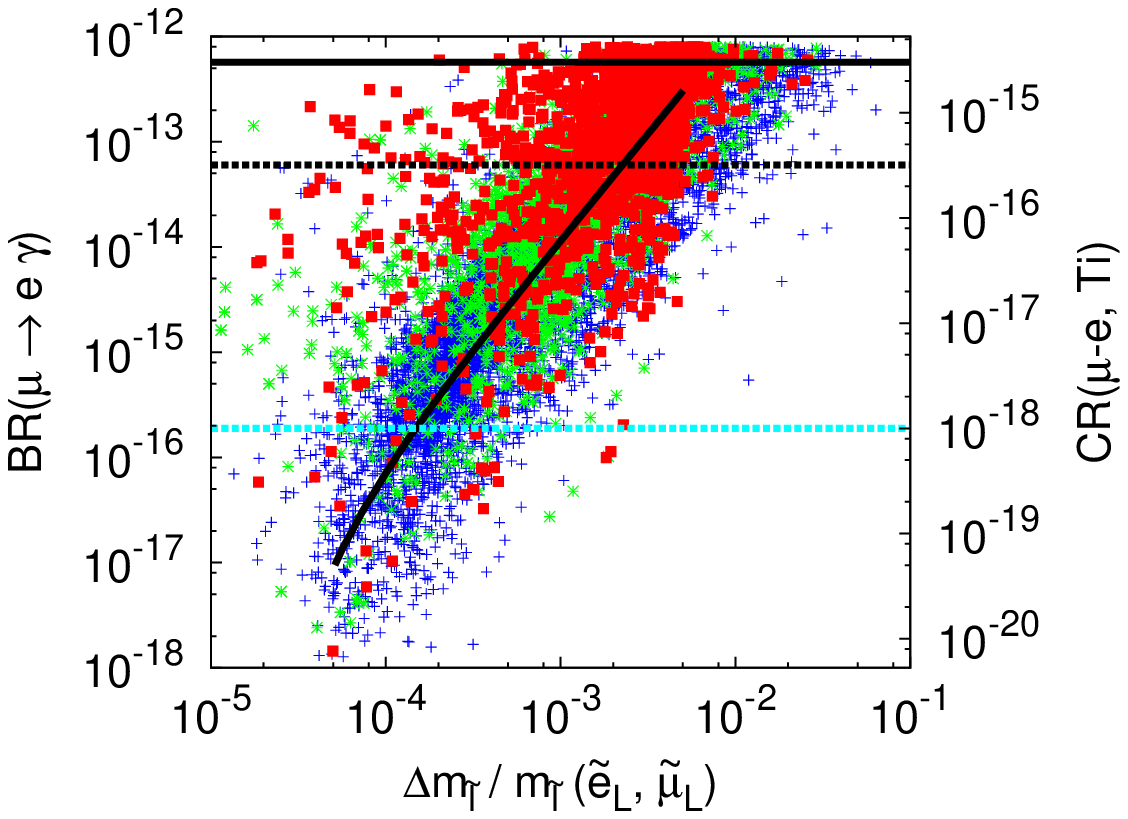, width=55mm}
\end{tabular}
\caption{\small 
On the left, example of invariant dimuon mass distributions from 
$\chi^0_2$ decays, for the cMSSM (dashed) and the SUSY seesaw (full
  lines), from~\cite{Abada:2010kj}. Central panel:
slepton mass splittings vs. average mass for a semi-constrained SUSY
spectrum.
We have taken $R = 1$, fixing the RH spectrum 
as $M_{R_1} = 10^{10}$ GeV, $M_{R_2} =
10^{11}$ GeV, and $M_{R_3} = 10^{12 - 15}$ GeV. Colour code
denotes regimes of BR($\mu \to e \gamma$): black points are already
excluded, and light green are within MEG reach, from~\cite{Figueiredo:2013tea}. 
On the right, BR($\mu \to e \gamma$) vs. slepton mass splittings.
Horizontal lines denote current
bounds and future sensitivity. The matrix $R$ is randomly varied
and the RH spectrum set as $M_{R_1} = 10^{10}$ GeV, $M_{R_2} =
10^{11}$ GeV, and $M_{R_3} = 10^{13, 14,15}$ GeV (blue, green and red
points), from~\cite{Figueiredo:2013tea}.}\label{fig:LHC1}
\end{figure}

In the cMSSM (i.e., in the absence of a seesaw mechanism), $\chi_2^0$
decays lead to identical flavour, opposite-sign, final state
leptons. The dielectron and dimuon invariant mass distributions
($m_{ee}$ and $m_{\mu \mu}$) have a double triangular shape, each
exhibiting two edges, which correspond to the exchange of the left-
and right-handed selectron or smuon. This is illustrated by the 
the dashed
lines on the left pannel of Fig.~\ref{fig:LHC1}, for two cMSSM
benchmark points (for details, see~\cite{Abada:2010kj}). 
A comparision of $m_{ee}$ and $m_{\mu \mu}$
would reveal that the edges are superimposed, 
implying that sleptons of the first two families
are degenerate in mass (to a very good
approximation): $m_{\tilde e_L (\tilde e_R)} \approx m_{\tilde \mu_L
  (\tilde \mu_R)}$. 
Should a type I SUSY seesaw be at work, then several differences would
be manifest in the dilepton mass distributions. Firstly, 
and as can be seen from inspection of  the full 
lines of the left pannel of Fig.~\ref{fig:LHC1}, 
a new
edge appears, corresponding to the presence of an additional state in
the   $\chi_2^0 \to \mu \mu \chi_1^0$ decays - other than intermediate
left- and right-handed smuons, a stau ($\tilde \tau_2$) also
contributes, $\chi_2^0 \to \tilde \tau_2 \mu \to\mu \mu
\chi_1^0$. This provides an indisputable {\it signal of cLFV at the LHC}.
Secondly, when comparing $m_{ee}$ and $m_{\mu \mu}$, the edges
corresponding to the left-handed (LH) 
sleptons are no longer superimposed, which
implies that left-handed sleptons of the first two families are no
longer degenerate, in other words, there will be a non-negligible
flavour violating slepton mass splitting, parametrised as 
\begin{equation}
\frac{\Delta m_{\tilde \ell}}{m_{\tilde \ell}} (\tilde e_l, \tilde \mu_L)
\, =\, \frac{|m_{\tilde e_L} -m_{\tilde \mu_L}|}{<m_{\tilde e_L},m_{\tilde \mu_L}>}\,.
\end{equation}
We now proceed to discuss the prospects for studying this observable (and
its synergy with low-energy cLFV observables) at the LHC, following
the most recent experimental results (in particular LHC SUSY negative
searches and the measurement of the Higgs mass, $m_h$, the recent  
MEG bound on $\mu \to e \gamma$ decays, and the determination
of $\theta_{13}$)~\cite{Figueiredo:2013tea}.
Our analysis 
has shown that in the framework of a type I
seesaw embedded in the fully constrained SUSY models, 
LHC data (in particular 
the measurement of $m_h$) precludes the possibility of 
simultaneously having BR($\mu \to e \gamma$)
within MEG reach and sizeable slepton mass differences 
associated with a slepton spectrum sufficiently light to be produced.
On the other hand, considering semi-constrained scenarios
allows to circumvent some of the strongest LHC bounds, 
especially on $m_h$, allowing for non-negligible slepton mass
splittings (for a comparatively light slepton spectrum). 
Depending on the SUSY regime, one can still have $\Delta m_{\tilde
  \ell} \sim 0.1\% - 1\%$, for 
$m_{\tilde  \ell}$ ranging from 800 GeV to 1.6 TeV.
This is illustrated on the central panel of Fig.~\ref{fig:LHC1}, which
also reveals that the associated rates for low-energy cLFV observables,
such as BR($\mu \to e \gamma$) and CR($\mu- e$, Ti) are also within 
experimental reach. 

Finally, the potential measurement of low- and high-energy cLFV
observables allows to probe the type I SUSY seesaw hypothesis
(for a type III SUSY seesaw, see~\cite{Abada:2011mg}): 
as shown on the right panel of Fig.~\ref{fig:LHC1},
these observables
exhibit a strong degree of correlation, even when fully exploring the
degrees of freedom of the neutrino Yukawa couplings (see
parametrisation of $Y^\nu$ given in Section~\ref{susyseesaw}). 
Any measurement $\Delta m_{\tilde \ell} \gtrsim \mathcal{O}(1\%)$
must be accompanied by the observation of $\mu \to e \gamma$
at MEG to substantiate the SUSY seesaw hypothesis; on the other hand, 
isolated manifestation of either low- or high-energy cLFV would
strongly suggest that sources of LFV, other than - or in addition to -
the SUSY seesaw would be present. 

\section{cLFV at a Linear Collider}
If SUSY is realised in Nature, 
Linear Colliders are ideal laboratories to study the slepton sector, 
offering a perfect environment
for high-energy precision studies such as cLFV in slepton production
and decays. The comparatively clean environment (due to a reduced QCD
activity), associated to the high-efficiency of the muon and electron
detectors implies that cLFV observables can be measured to a very good
precision; beam polarisation
allows to reduce the SM backgrounds and probe the chirality structure
of the new processes, and the possibility of having an $e^-e^-$ beam
option offers a powerful tool to further probe the properties of the
new states.
Provided that the centre of mass energy is sufficiently high, one can
have direct slepton production, and study slepton phenomena in
short(er) SUSY decay chains. 
 
In addition to the observables already considered in the previous section
devoted to the LHC (i.e. edges in dilepton mass distributions, slepton
mass splittings), there are many other cLFV phenomena that can be
probed at a LC. Here we focus on $e^\pm e^- \to e^\pm \mu^- +
E_\mathrm{miss}$, where the missing energy can be in the form of
lightest neutralinos (cLFV signal), neutralinos and neutrinos
(corresponding to the SUSY charged current background - flavour
violating vertices due to the presence of the $U_\mathrm{PMNS}$) and 
neutrinos (SM background, again due to $U_\mathrm{PMNS}$ flavour
mixing in charged currents. We will also discuss the potential 
for cLFV of $e^- e^- \to \mu^- \mu^- +E_\mathrm{miss}$.
A detailed discussion of the different processes and observables 
can be found in~\cite{Abada:2012re}.

\subsection{$\pmb{e^+ e^-}$ beam option}
We begin with an illustration of the potential of $e^+ e^-$ beam
option for cLFV: on the left panel of Fig.~\ref{fig:LC1} 
we display the cross section of different
processes contributing to $e^+ e^-
\to e^+ \mu^- +E_\mathrm{miss}$. We consider a sample benchmark point,
with conservative SUSY seesaw assumptions  (i.e., $R=1$, with 
all FV arising from the  $U_\mathrm{PMNS}$, and an intermediate RH neutrino
scale, $M_R \sim 10^{12}$ GeV). Relevant information 
on the neutralino and slepton
spectra is also summarised in Fig.~\ref{fig:LC1}. 

Although the SM$_{m_\nu}$ background clearly dominates, it is expected
to be easily disentangled from SUSY events via appropriate
cuts. Provided that $\sqrt s$ is above the slepton production
threshold, the cLFV signal dominates the MSSM$_{m_\nu}$ background. 
The left panel of Fig.~\ref{fig:LC1} 
also conveys the potential of beam polarisation: to fully reveal
it, we considered an ideal scenario where both beams are fully
polarised. As can be seen, polarising the beams reduces the
SM$_{m_\nu}$ background and essentially removes the SUSY one.

\subsection{$\pmb{e^- e^-}$ beam option}
In the absence of doubly charged particles (as
is the case of the SUSY seesaw), all slepton production in 
$e^- e^-$ collisions occurs via
t-channel neutralino exchange. Having similar slepton production for
signal and SUSY background implies that polarising the beams will have
a smaller effect. 

\begin{figure}
\begin{tabular}{ccc}
 \epsfig{file=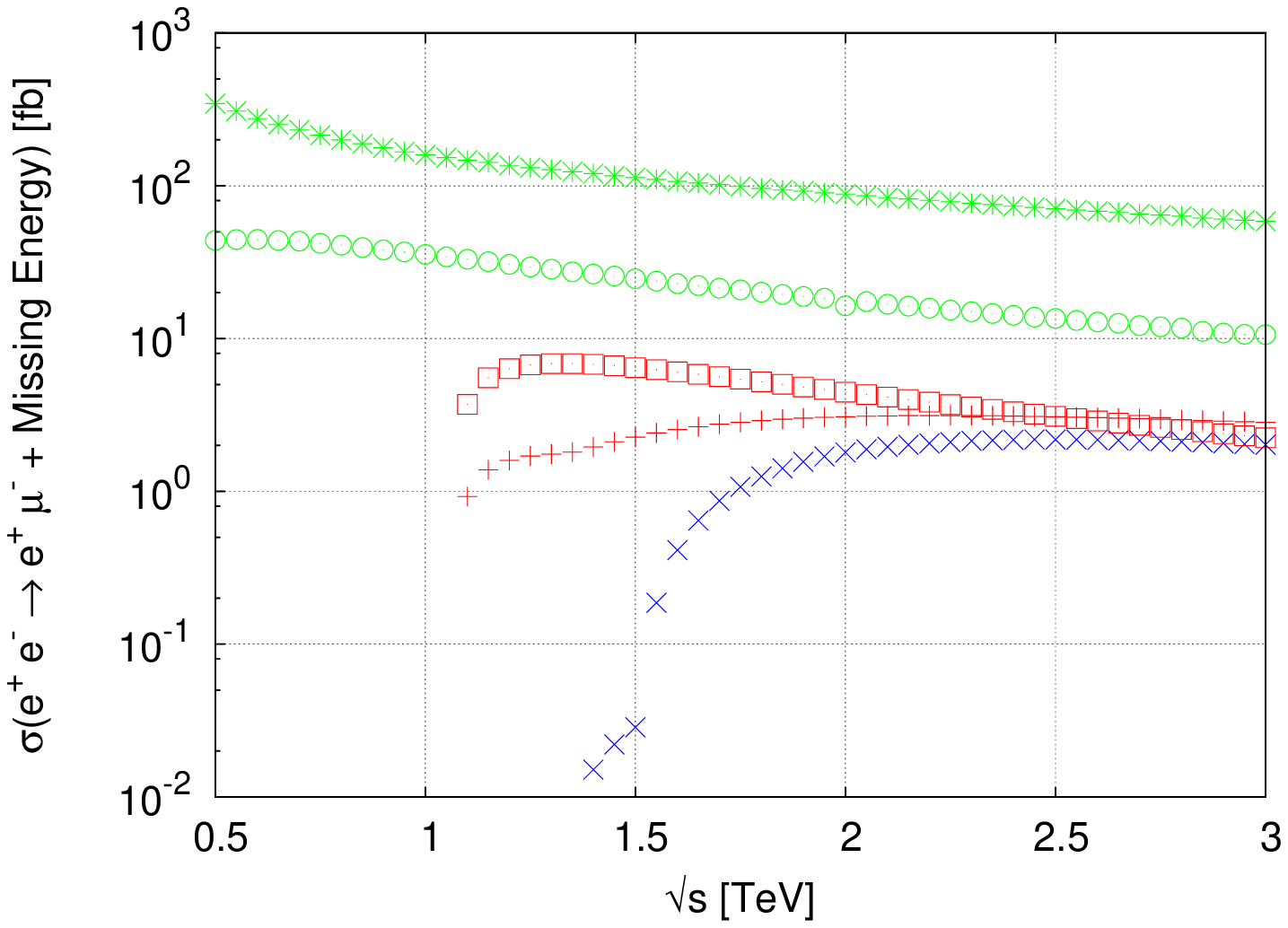,     clip=, angle=0, width=55mm}  
&\hspace*{5mm}
\raisebox{36mm}{\epsfig{file=figs/table.spectrum.epsi,width=32mm,angle=270,clip=}}
\hspace*{5mm}&
\epsfig{file=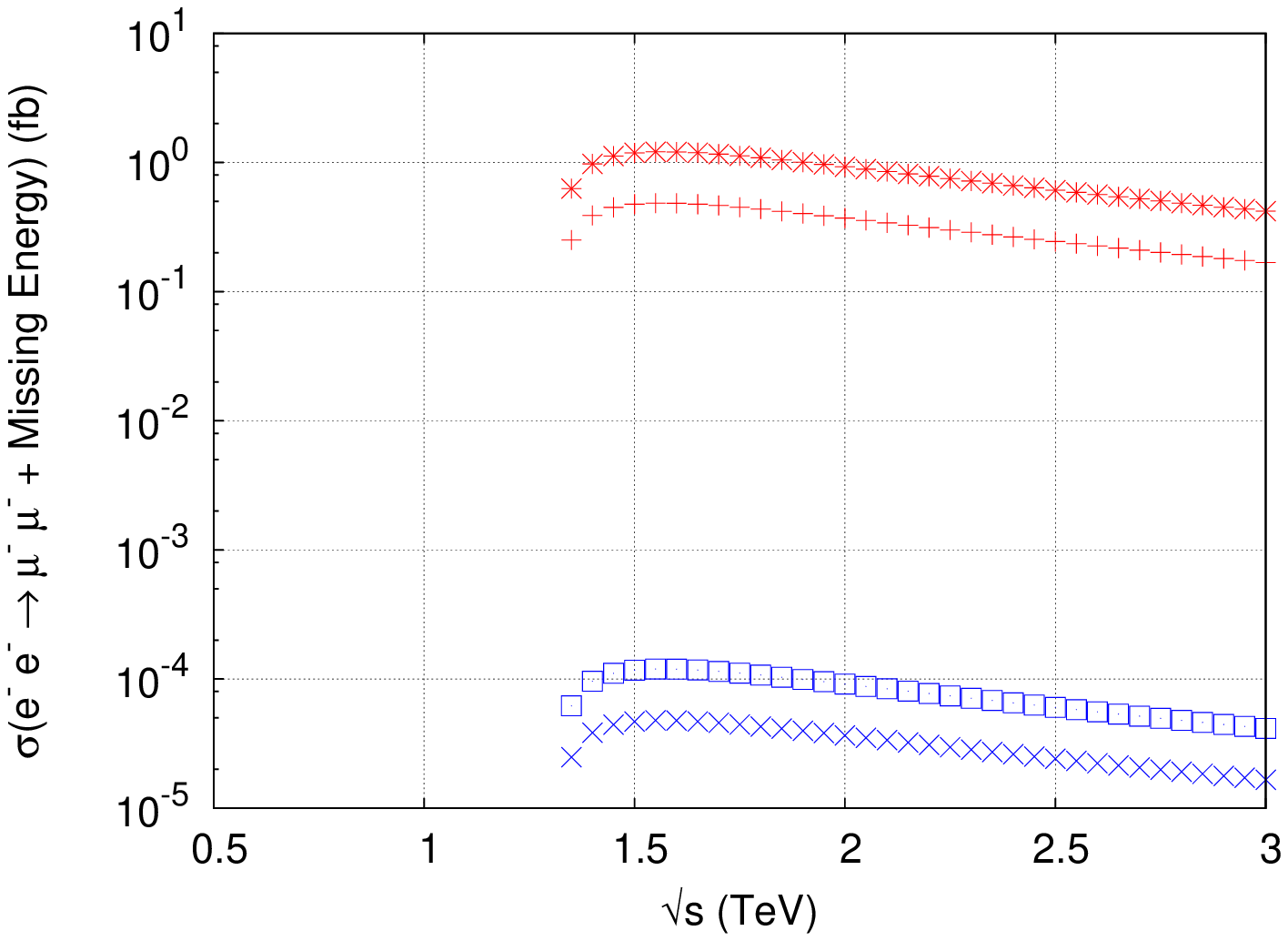,width=55mm,angle=0,clip=}
\end{tabular}
\caption{\small On the left, $\sigma(e^+ e^- \to e^+
\mu^- + E^T_\text{miss})$, {with $E^T_\text{miss}=2 \chi_1^0$ (signal,
  red), $2 \chi_1^0
+  (2,4)\nu$ (SUSY background, blue) and  $(2,4) \nu$ (SM background,
green)}, as a function of the  centre of mass energy,
$\sqrt{s}$. We fix $M_R = 10^{12}$ GeV, for a SUSY spectrum 
summarised by the Table above, showing the results for 
both unpolarised (crosses, times
and asterisks) and fully
polarised beams (squares and circles). 
On the right, $\sigma(e^- e^- \to \mu^-
\mu^- + E^T_\text{miss})$, for the same SUSY seesaw choice, also in
the unpolarised and fully polarised case. From~\cite{Abada:2012re}.
\label{fig:LC1}}
\end{figure}

As mentioned before, the $e^- e^-$ beam option is ideal for what might
be a truly ``golden channel'' of cLFV at a LC: $e^- e^- \to \mu^- \mu^-
+ E_\mathrm{miss}$.
When compared to other signals of cLFV at a LC already discussed -
e.g. $e^{\pm} e^- \to e^{\pm} \mu^- +
E^T_\text{miss}$ - the SM model background is extremely
tiny in this case. SUSY background processes are still present, but are
subdominant when compared to the signal, as the corresponding cross
sections differ by at least one order of magnitude. This is
illustrated in the right panel of 
Fig.~\ref{fig:LC1}, for the same points previously considered.
Notice that the signal clearly dominates (by about four orders of
magnitude) the SUSY background.

Provided that the c.o.m. energy is sufficiently large (above
the slepton production threshold), $e^{-} e^- \to \mu^- \mu^- +
E^T_\text{miss}$ is an ideal cLFV discovery channel, allowing 
the identification of a new
physics scenario, such as the SUSY seesaw - always under the 
assumption of having a unique source of Lepton Flavour Violation
present. Naturally, one would be also confirming the Majorana nature
of the exchanged particles in the t-channel.
Moreover, right-polarising both beams would allow to test the seesaw
hypothesis since, in such a framework, slepton cLFV is predominantly a
left-sector phenomenon.

Although not discussed here, one can also study the synergy between
low-energy cLFV observables and high-energy ones measured
at a LC (similar to the LHC)~\cite{Abada:2012re}. 
In addition to probing the hypothesis of
a type I SUSY seesaw, such a study can also hint towards the seesaw scale.

\section{Outlook}
Lepton flavour violation observables play a leading r\^ole in
unveilying the presence of NP. Here we have discussed the potential of
high-energy colliders (LHC and a future LC) for the study of cLFV
signals. As an example, we have considered the appealing case of a type
I SUSY seesaw. 

At the LHC, the most striking cLFV signal would be the appearance of
multiple edges in dimuon mass distributions from EW-ino decays. We
have also considered the prospects for observing flavoured slepton
mass differences (and exploring their synergy with low-energy cLFV
observables) finding that semi-constrained SUSY scenarios still offer
promising prospects for these studies. 

We have revisited the case for cLFV at a Linear Collider, exploring
both beam options and beam polarisation, further discussing the
relevant backgrounds for charged current processes. Provided that the
c.o.m. energy is above the slepton production threshold, one can
indeed expect abundant events (even prior to selection cuts). We have
also pointed a possibly ``golden channel'' for cLFV, 
$e^- e^- \to \mu^-\mu^- + E_\text{miss}$, one 
of the cleanest experimental setups to probe both cLFV and  lepton
number violation. 

\section*{Acknowledgments}
We acknowledge partial support from the European Union FP7 ITN
INVISIBLES (Marie Curie Actions, PITN-GA-2011-289442), and from 
the ``HADRONPHYSICS3'' contract.

\end{document}